\documentclass[twocolumn,prl,showpacs,superscriptaddress]{revtex4}

\usepackage{graphicx}
\usepackage{dcolumn}
\usepackage{bm}

\usepackage{amsmath, relsize}

\begin{document}

\title{A Hafnian PH-Pfaffian State for $\nu = 5/2$ Quantum Hall Effect}
\author{Jian Yang}
\email{jyangmay1@yahoo.com}
\altaffiliation{Permanent address: 4610 Ravensthorpe Ct, Sugar Land, TX 77479, USA}
%\affiliation{}
%\date{}

\begin{abstract}

The PH-Pfaffian state having $1/2$ central charge is consistent with the thermal Hall conductance measurement of $\nu = 5/2$ fractional quantum Hall system, but lacks support from the existing numerical results. In this paper we propose a new state described by a wavefunction obtained by multiplying ${Haf} ( z_i^*-z_j^* )^2 $ to a PH-Pfaffian wavefunction, with $Haf(A)$ being the Hafnian of a symmetric matrix $A$. We call this new state the Hafnian PH-Pfaffian state. In spherical geometry, the Hafnian PH-Pfaffian state has the same magnetic flux number $N_{\phi}= 2N-3$ as the Pfaffian state, allowing a direct numerical comparison between the two states. Results of exact diagonalization of finite systems in the second Landau level show that the overlap of the exact ground state with the Hafnian PH-Pfaffian state exceeds that with the Pfaffian state when the short range component of the Coulomb interaction increases to a certain level, lending a numerical support to the Hafnian PH-Pfaffian state.
We further show the Hafnian PH-Pfaffian state is mathematically identical to the newly proposed compressed PH-Pfaffian state [arXiv:2001.01915 (2020)] formed by ”compressing” the PH-Pfaffian state with two flux quanta removed to create two abelian Laughlin type quasiparticles of the maximum avoidance from one another. As the result we argue that the Hafnian PH-Pfaffian state has the same central charge as the PH-Pfaffian state, and is therefore consistent with the thermal Hall conductance measurement. Finally we present numerical results on two new wavefunctions formed by increasing the relative angular momentum by two for each of the paired composite fermions of the PH-Pfaffian state.
\end{abstract}
\pacs{73.43.Cd, 71.10.Pm } \maketitle

After thirty five years since its discovery\cite{Willett}, the understanding of the fractional quantum Hall effect (FQHE) at $\nu=5/2$ filling factor is still elusive and remains a great intellectual challenge. Among the three primary candidates, Pfaffian state\cite{MR} and its PH conjugate, the anti-Pfaffian state\cite{Levin} \cite{Lee}, form two degenerate but distinct states in the absence of Landau level mixing. When the Landau level mixing is properly taken into account, the degeneracy between the Pfaffian and anti-Pfaffian states is lifted, and the numerical studies are in favor of the anti-Pfaffian state energetically\cite{EHRezayi}, making the anti-Pfaffian state a more likely a candidate for the ground state of the $\nu=5/2$ FQHE. The third topologically different state that is particle-hole (PH) symmetric, hence termed PH-Pfaffian state \cite{Son}\cite{Zucker}\cite{Yang} has also attracted a great attention. Unfortunately, it lacks any numerical support. In spherical geometry, no consistent gapped ground state is found to exist at the total flux number $N_{\phi} = 2N-1$ as required by the PH symmetry even with a wide range of variations of Coulomb interactions\cite{Yang}. This is consistent with the findings that the PH-Pfaffian state may fail to represent a gapped, incompressible phase\cite{Mishmash}\cite{Balram}.

While the existing numerical results seem to converge to a consensus  that the anti-Pfaffian state is the most likely candidate for the ground state of the $\nu=5/2$ FQHE, the thermal Hall conductance measurement\cite{Banerjee} casts a great doubt on this consensus. Although the thermal Hall conductance measurement, $\kappa_{xy} = 5/2$ in units of $\frac{\pi^2k^2_B}{3h}T$, is rather encouraging and pointing to the existence of non-abelian quasiparticles, it is incompatible with the edge structure of anti-Pfaffian, but rather consistent with the PH-Pfaffian topological order. In view of the discrepancy between the numerical and experimental results, some more complicated proposals are put forward to resolve the discrepancy. Among them are disorder induced mesoscopic puddles composed of Pfaffian and anti-Pfaffian states\cite{Mross}\cite{CWang}\cite{Lian}, and incomplete thermal equilibration on an anti-Pfaffian edge\cite{Simon}\cite{Ma1}\cite{Asasi}\cite{Rosenow}, which are all under debate \cite{Feldman}.

In the spherical geometry, the Pfaffian state is formed at $N_{\phi}= 2N-3$, the anti-Pfaffian at $N_{\phi}= 2N+1$, and the PH-Pfaffian state at $N_{\phi}= 2N-1$. In an attempt to search for numerical support for the PH-Pfaffian topological order, in \cite{YangCompressed} we asked the following question: can we form an incompressible state at $N_{\phi}= 2N-3$ or $N_{\phi}= 2N+1$ that is not PH symmetric yet maintains the PH-Pfaffian topological order and is energetically more favorable (or has larger overlap with the exact ground state) than the Pfaffian state or anti-Pfaffian state, at least for a certain parameter range of the Coulomb interactions? The answer was yes. The idea is to add two Laughlin type abelian quasiparticles to the PH-Pfaffian state, and make the two quasiparticles form a uniform state with the maximum avoidance from one another. The resulting state is termed as a compressed PH-Pfaffian state, as it can be viewed as the result of "compressing" the PH-Pfaffian state with two flux quanta removed. The compressed PH-Pfaffian state is not PH symmetric but possesses the PH-Pfaffian topological order. Since both the compressed PH-Pfaffian state and the Pfaffian state formed at $N_{\phi} = 2N-3$, it allows for a direct numerical comparison between the two. The finite size numerical results show that the overlap of the exact ground state with the compressed PH-Pfaffian state exceeds that with the Pfaffian state when the short range component of the Coulomb interaction increases to a certain level, lending a numerical support to the compressed PH-Pfaffian state.

While the compressed PH-Pfaffian state blows a life to the PH-Pfaffian topological order numerically, a question was raised if the finite size result will survive in the thermal dynamic limit as one would think two quasiparticles in the ground state would make no difference in the thermal dynamic limit \cite{FeldmanPrivate}\cite{HalperinPrivate}. It is the main purpose of this brief report to address this apparent "two quasiparticles" thermal dynamic shortcoming with a more elegant resolution. To this end, we propose a new state described by the following wave function 
\begin{equation}
\label{HPH-Pfaffian} {\Psi}_{HPH} = {Haf} ( z_i^*-z_j^* ) ^2  {\Psi}_{PH}
\end{equation}
where $z_j = x_j+iy_j$ is the complex coordinate of the $j_{th}$ electron, $z_j^* = x_j-iy_j$, $N$ is the total number of electrons,  $Haf(A)$ is the Hafnian of a $N$ by $N$ symmetric matrix $A$ with $N$ being an even integer, 
\begin{equation}
\label{Hafnian}{Haf} (A) =\mathlarger{\sum}_{\sigma \in S_{N}} \prod\limits_{j=1}^{N/2} A_{\sigma(2j-1),\sigma(2j)}
\end{equation}
where $S_{N}$ is the symmetric group on $[N] = {1, 2, \cdot\cdot\cdot, N}$, and
\begin{equation}
\label{PH-Pfaffian} {\Psi}_{PH} = {Pf} ( \frac{1}{z_i^*-z_j^* } )   \prod\limits_{i<j}^N (z_i-z_j)^2
\end{equation}
is a PH-Pfaffian wave function\cite{Zucker}. We call this new state described Eq.(\ref{HPH-Pfaffian}) the Hafnian PH-Pfaffian state, and obviously it is not subject to the apparent "two quasiparticles" thermal dynamic shortcoming suffered by the compressed PH-Pfaffian state as discussed above.  

To make it numerically easier to deal with when projecting to the lowest Landau level\cite{Yang}, we will use an alternative form of Eq.(\ref{HPH-Pfaffian}) and Eq.(\ref{PH-Pfaffian}):
\begin{eqnarray}
\label{HPH-Pfaffian-A}
{\Psi}_{HPH} = {Haf}( \frac{\partial}{{\partial}z_i}-\frac{\partial}{{\partial}z_j} )^2 {\Psi}_{PH}
\end{eqnarray}
\begin{eqnarray}
\label{PH-Pfaffian-A}
{\Psi}_{PH} = {Pf}(\frac{1}{ \frac{\partial}{{\partial}z_i}-\frac{\partial}{{\partial}z_j} })\prod\limits_{i<j}^N(\frac{\partial}{{\partial}z_i}-\frac{\partial}{{\partial}z_j} )  \prod\limits_{i<j}^N (z_i-z_j)^3
\end{eqnarray}
or written in spherical geometry
\begin{eqnarray}
\label{SHPH-Pfaffian} &&
{\Psi}_{HPH} = {Haf} (\frac{\partial}{{\partial}u_i}\frac{\partial}{{\partial}v_j}-\frac{\partial}{{\partial}u_j}\frac{\partial}{{\partial}v_i} )^2 {\Psi}_{PH}
\end{eqnarray}
\begin{eqnarray}
\label{SPH-Pfaffian}
{\Psi}_{PH} = {Pf}(\frac{ 1}{  \frac{\partial}{{\partial}u_i}\frac{\partial}{{\partial}v_j}-\frac{\partial}{{\partial}u_j}\frac{\partial}{{\partial}v_i} })  
\prod\limits_{i<j}^N(\frac{\partial}{{\partial}u_i}\frac{\partial}{{\partial}v_j}-\frac{\partial}{{\partial}u_j}\frac{\partial}{{\partial}v_i} ) \Phi_{3} 
\end{eqnarray}
where $ \Phi_{3} = \prod\limits_{i<j}^N (u_iv_j-u_jv_i)^{3}$  and $(u, v)$ are the spinor variables describing electron coordinates.

Since the total flux number $N_{\phi}$ corresponding to the PH-Pfaffian wave function is $N_{\phi} = 2N-1$,  the Hafnian PH-Pfaffian has $N_{\phi} = 2N-3$. This is the same $N_{\phi}-N$ relationship for the Pfaffian state.
The fact that both  the Hafnian PH-Pfaffian state and the Pfaffian state have the same $N_{\phi}-N$ relationship $N_{\phi} = 2N-3$, allows for a direct numerical comparison between the two to determine which wave function, therefore which topological order and at what condition, represents the exact ground state. In Fig.1, we calculated and plotted the overlap of the exact ground state of a finite system ($N_{\phi}, N$) = ($13, 8$) with the Pfaffian state and with the Hafnian PH-Pfaffian state respectively. The exact ground state is obtained in the second Landau level free of Landau level mixing, with  the  ratios of $V_1/V_1^c$ ranging from $1$ to $1.5$, where $V_1^c$ is the Coulomb value of $V_1$ in the second Landau level. We see the ground state undergoes a phase transition from the Pfaffian state to the Hafnian PH-Pfaffian state as the short range component of the Coulomb interaction increases. The transition occurs at $V_1/V_1^c$ around $1.2$, as at this point the overlap of the exact ground state with the Hafnian PH-Pfaffian state exceeds that with the Pfaffian state. 
 
\begin{figure}[tbhp]
\label{fig:Overlap}
%\vspace{0.2cm}
%\includegraphics[width=18cm,height=12cm]{Overlap_Pfaffian_Hafnian1_8e}
\includegraphics[width=\columnwidth]{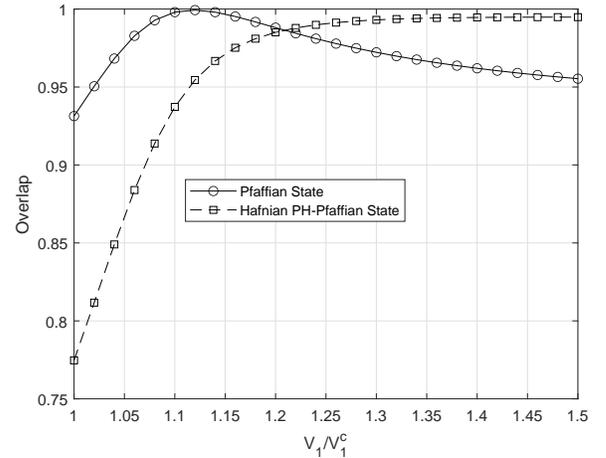}
\caption{\label{fig:Overlap} For $N=8$ and $N_{\phi} = 13$. Overlap of the exact ground state with the Pfaffian state (solid line) and the 
Hafnian PH-Pfaffian state (dashed line) as the function of the pseudopotential $V_1$
normalized by its Coulomb value $V_1^c$ in the second Landau level. The overlap between the Pfaffian state and the 
Hafnian PH-Pfaffian state is $0.9505$.}
\end{figure}

If we compare the Fig.1 in this paper with the Fig.1 in \cite{YangCompressed}, they look extremely similar. In fact, we find the overlap between the Hafnian PH-Pfaffian wavefunction Eq.(\ref{SHPH-Pfaffian}) and the compressed PH-Pfaffian wavefunction is exactly one. This is rather surprising as the Hafnian wavefunction Eq.(\ref{SHPH-Pfaffian}) looks very different from the following compressed PH-Pfaffian wavefunction written on a sphere:
\begin{equation}
\label{SCPH-Pfaffian}
\int d{\Omega}_1 d{\Omega}_2 (\alpha_1\beta_2-\alpha_2\beta_1)^N \prod\limits_{i=1}^N \prod\limits_{a=1}^2
(\beta_a^*\frac{\partial}{{\partial}u_i}-\alpha_a^*\frac{\partial}{{\partial}v_i}){\Psi}_{PH}
\end{equation}
 where $\prod\limits_{i=1}^N \prod\limits_{a=1}^2
(\beta_a^*\frac{\partial}{{\partial}u_i}-\alpha_a^*\frac{\partial}{{\partial}v_i})$ creates two Laughlin type abelian quasiparticles located at the spinor variables $({\alpha}_1, {\beta}_1)$ and $({\alpha}_2, {\beta}_2)$  from the PH-Pfaffian state, and the two quasiparticles form a uniform state with the maximum avoidance from one another (or the maximum number, $N$, of zeros) in the form of $ (\alpha_1\beta_2-\alpha_2\beta_1)^N$.  One can carry out the integration over the quasiparticles coordinates and rewrite Eq.(\ref{SCPH-Pfaffian}) as:
\begin{equation}
\label{SCPH-Pfaffian1} {\Psi}_{CPH} = \sum\limits_{m=-\frac{N}{2}}^{\frac{N}{2}}  (-1)^m G_{m}G_{-m} {\Psi}_{PH}
\end{equation}
where 
\begin{eqnarray}
&&G_{m}=(-1)^{\frac{N}{2}-m}[\frac{N!}{(\frac{N}{2}+m)!(\frac{N}{2}-m)!}]^{-1/2}\cdot
 \nonumber \\
&&
\sum_{1{\leq}l_1<l_2<{\ldots}{\leq}l_{\frac{N}{2}+m}}\frac{\partial}{\partial
v_{l_1}}\frac{\partial}{\partial
v_{l_2}}\ldots\frac{\partial}{\partial v_{l_{\frac{N}{2}+m}}} \cdot\nonumber \\
&& \prod_{l({\neq} l_1, l_2, \ldots, l_{\frac{N}{2}+m})}
\frac{\partial}{\partial u_l}.
\end{eqnarray}
is a quasiparticle generation operator in angular momentum space \cite{YangHierarchy}, which
will generate a quasiparticle with angular momentum $(L,L_z)=(\frac{N}{2},m)$ when applied to ${\Psi}_{PH}$ which has angular momentum $L=0$. As a result, ${\Psi}_{CPH}$ is formed from the two quasiparticles  and has total angular momentum $L=0$, thus is rotationally invariant, a condition required for an incompressible state in the spherical geometry. 

The reason for the perfect unity overlap between the Hafnian PH-Pfaffian wavefunction Eq.(\ref{SHPH-Pfaffian}) and the compressed PH-Pfaffian wavefunction Eq.(\ref{SCPH-Pfaffian1}) is traced back to the following mathematical identity which we have validated numerically:
\begin{eqnarray}
&&
Haf(a_i-a_j)^2 = 
\nonumber \\
&&
\mathlarger{\sum}_{k = -\frac{N}{2}}^{\frac{N}{2}} (-1)^k \frac{(\frac{N}{2}+k)!(\frac{N}{2}-k)!}{(\frac{N}{2})!} e_{\frac{N}{2}+k}e_{\frac{N}{2}-k}
\end{eqnarray}
where $e_k$ is the fundamental symmetric polynomials  of variables $a_i (i = 1, 2, \cdot\cdot\cdot, N)$ with $N$ being an even integer
\begin{eqnarray}
e_k = \mathlarger{\sum}_{1\leq i_1<i_2< \cdot\cdot\cdot <i_k \leq N} a_{i_1}a_{i_2}\cdot\cdot\cdot a_{i_k}
\end{eqnarray}
The fact that the Hafnian PH-Pfaffian $Haf(a_i-a_j)^2$ can be written as the sum of the products of two fundamental symmetric polynomials  demonstrates that the Hafnian PH-Pfaffian state has the same central charge as the PH-Pfaffian state, and is therefore consistent with the thermal Hall conductance measurement. As there exists a PH conjugate state of the Pfaffian state, the anti-Pfaffian, there also exists a PH conjugate state of the Hafnian PH-Pfaffian state. Since the number of
electrons $N$ is related to the number  of holes $N_h$ of the PH conjugate state by $N+N_h = N_{\phi}+1$, the relationship between the flux number and the number of holes of the anti-Pfaffian or the PH conjugate state of the Hafnian PH-Pfaffian state is $N_{\phi} = 2N_h+1$. While Pfaffian and anti-Pfaffian are two topologically distinct states, we believe the Hafnian PH-Pfaffian state and its PH conjugate state have the same topological order. In the absence of PH symmetry breaking factors such as Landau level mixing, the same transition from the anti-Pfaffian state to the PH conjugate of the Hafnian PH-Pfaffian state will take place at the same short range interaction strength.

Similar to Eq.(\ref{HPH-Pfaffian}), we can form another Hafnian wavefunction
\begin{equation}
\label{HPH-Pfaffian1} {Haf} ( z_i-z_j ) ^2  {\Psi}_{PH}
\end{equation}
As expected, this wavefunction is mathematically identical to the "stretched" PH-Pfaffian wavefunction in \cite{YangCompressed}. It occurs at the same flux number as the anti-Pfaffian state and the overlap between the two is $0.7187$. One can switch the order between the Hafnian factor ${Haf} ( z_i-z_j ) ^2$ in Eq.(\ref{HPH-Pfaffian1}) and the Pfaffian factor ${Pf} ( \frac{1}{z_i^*-z_j^* } ) $ in ${\Psi}_{PH}$ to form a different wavefunction which has a larger overlap $0.8329$ with the anti-Pfaffian state.

Finally, we would like to present numerical results for the following two wavefunctions:
\begin{equation}
\label{IPH-Pfaffian-1} {Pf} ( \frac{(z_i-z_j)^2}{z_i^*-z_j^* } )    \prod\limits_{i<j}^N (z_i-z_j)^2
\end{equation}
and
\begin{equation}
\label{IPH-Pfaffian-2} {Pf} ( \frac{z_i-z_j}{(z_i^*-z_j^*)^2 } )    \prod\limits_{i<j}^N (z_i-z_j)^2
\end{equation}
which can be viewed as the result of increasing the relative angular momentum by two for each of the paired composite fermions of the PH-Pfaffian state. 

Again, to make it numerically easier to deal with when projecting to the lowest Landau level\cite{Yang}, we will use an alternative form of Eq.(\ref{IPH-Pfaffian-1}):
\begin{eqnarray}
\label{IPH-Pfaffian-1A}
{Pf}(\frac{(z_i-z_j)^2}{ \frac{\partial}{{\partial}z_i}-\frac{\partial}{{\partial}z_j} })
\prod\limits_{i<j}^N(\frac{\partial}{{\partial}z_i}-\frac{\partial}{{\partial}z_j} )  \prod\limits_{i<j}^N (z_i-z_j)^3
\end{eqnarray}
or written in spherical geometry
\begin{equation}
\label{SIPH-Pfaffian-1A} {Pf}(\frac{ (u_iv_j-u_jv_i)^{2}}{  \frac{\partial}{{\partial}u_i}\frac{\partial}{{\partial}v_j}-\frac{\partial}{{\partial}u_j}\frac{\partial}{{\partial}v_i} })  \prod\limits_{i<j}^N(\frac{\partial}{{\partial}u_i}\frac{\partial}{{\partial}v_j}-\frac{\partial}{{\partial}u_j}\frac{\partial}{{\partial}v_i} ) {\Phi}_{3} 
\end{equation}
where ${\Phi}_{3} =   \prod\limits_{i<j}^N (u_iv_j-u_jv_i)^{3} $ is the Laughlin wave function. Similarly, we will use an alternative form of Eq.(\ref{IPH-Pfaffian-2}):
\begin{eqnarray}
\label{IPH-Pfaffian-2A}
{Pf}(\frac{z_i-z_j}{ (\frac{\partial}{{\partial}z_i}-\frac{\partial}{{\partial}z_j})^2 })
\prod\limits_{i<j}^N(\frac{\partial}{{\partial}z_i}-\frac{\partial}{{\partial}z_j} )  \prod\limits_{i<j}^N (z_i-z_j)^3
\end{eqnarray}
or written in spherical geometry
\begin{equation}
\label{SIPH-Pfaffian-2A} {Pf}(\frac{ u_iv_j-u_jv_i}{  (\frac{\partial}{{\partial}u_i}\frac{\partial}{{\partial}v_j}-\frac{\partial}{{\partial}u_j}\frac{\partial}{{\partial}v_i} )^2})  \prod\limits_{i<j}^N(\frac{\partial}{{\partial}u_i}\frac{\partial}{{\partial}v_j}-\frac{\partial}{{\partial}u_j}\frac{\partial}{{\partial}v_i} ) {\Phi}_{3} 
\end{equation}

Since the total flux number $N_{\phi}$ corresponding to the PH-Pfaffian wave function is $N_{\phi} = 2N-1$, and the wavefunctions Eq.(\ref{SIPH-Pfaffian-1A}) and Eq.(\ref{SIPH-Pfaffian-2A}) are formed from the PH-Pfaffian state by increasing the relative angular momentum by two for each of the paired composite fermions of the PH-Pfaffian state, the relationship between the flux number $N_{\phi}$ and the number of electrons $N$ is $N_{\phi} = 2N+1$. This is the same $N_{\phi}-N$ relationship for the anti-Pfaffian state. In Fig.2, we calculated and plotted the overlap of the exact ground state of a finite system ($N_{\phi}, N$) = ($13, 6$) with the anti-Pfaffian state and with the wavefunctions Eq.(\ref{SIPH-Pfaffian-1A}) and Eq.(\ref{SIPH-Pfaffian-2A}) respectively. 
\begin{figure}[tbhp]
\label{fig:Overlap1}
%\vspace{0.2cm}
%\includegraphics[width=18cm,height=12cm]{Overlap_Pfaffian_Hafnian2_8e}
\includegraphics[width=\columnwidth]{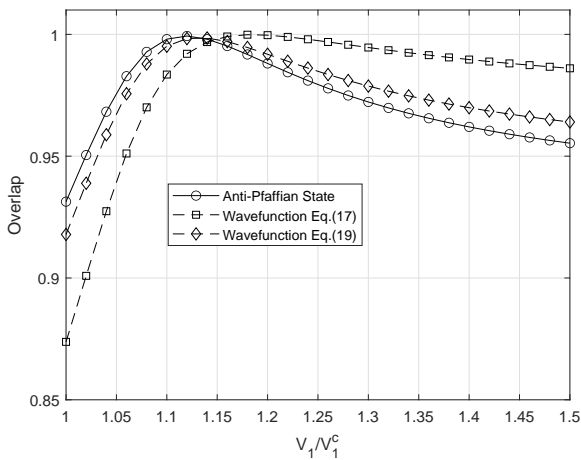}
\caption{\label{fig:Overlap} For $N=6$ and $N_{\phi} = 13$. Overlap of the exact ground state with the anti-Pfaffian state (solid line),  the 
wavefunction Eq.(\ref{SIPH-Pfaffian-1A}) (square-dashed line), and the wavefunction Eq.(\ref{SIPH-Pfaffian-2A}) (diamond-dashed line) as the function of the pseudopotential $V_1$
normalized by its Coulomb value $V_1^c$ in the second Landau level.}
\end{figure}
 
It should be pointed out that in the original version of this paper we claimed that the wavefunction Eq.(\ref{IPH-Pfaffian-1}) describes a state with the same topological order as the PH-Pfaffian state. This is incorrect since the numerator inside the Pfaffian symbol $(z_i-z_j)^2$ changes the non-abelian quasiparticle Hilbert space of the PH-Pfaffian state. This is in part the reason that leads to the proposal of the Hafnian PH-Pfaffian state by taking the numerator $(z_i-z_j)^2$ out of the Pfaffian and put it in the Hafnian. 

Furthermore, although it has a large overlap $0.9914$ with the anti-Pfaffian state, the wavefunction Eq.(\ref{SIPH-Pfaffian-1A}) has a repulsion rather than attraction between members of a pair, its behavior is not clear in a larger system. On the other hand, the overlap of the anti-Pfaffian state with wavefunction Eq.(\ref{SIPH-Pfaffian-2A}) is $0.9993$, and the wavefunction has an attraction between members of a pair, it is likely to have the same topological order as the anti-Pfaffian state\cite{HalperinPrivate1}.

Before closing, we would like to point out that Eq.(\ref{IPH-Pfaffian-1}) can be generalized to:
\begin{equation}
\label{IPH-Pfaffian_G1} {Pf} ( \frac{(z_i-z_j)^p}{(z_i^*-z_j^*)^q } )    \prod\limits_{i<j}^N (z_i-z_j)^m
\end{equation}
and
\begin{equation}
\label{IPH-Pfaffian_G2} {Pf} ( \frac{(z_i^*-z_j^*)^p}{(z_i-z_j)^q } )    \prod\limits_{i<j}^N (z_i-z_j)^m
\end{equation}
where $p$, $q$, and $m$ are positive integers, with $m \geq q$ and $m+p+q$ being odd for fermions and even for bosons. 

More works are required on a few fronts: First we need to see if the parameter range of the Coulomb interaction at which the system is in the Hafnian PH-Pfaffian state matches realistic conditions. Secondly, we need to study larger size finite systems to verify if the Hafnian PH-Pfaffian state can survive in the thermal dynamic limit.

\end{document}